\documentclass[%
 aip,
 amsmath,amssymb,
 reprint,%
]{revtex4-1}
\usepackage{graphicx}
\usepackage{dcolumn}
\usepackage{bm}

\usepackage[utf8]{inputenc}
\usepackage[T1]{fontenc}
\usepackage{mathptmx}
\usepackage{etoolbox}

\usepackage{xcolor}
\usepackage[separate-uncertainty = true,multi-part-units=single]{siunitx}

\usepackage{braket}

\usepackage[acronym]{glossaries}

\newacronym{qkd}{QKD}{Quantum Key Distribution}

\newacronym[plural=SNSPDs]{sns}{SNSPD}{superconducting nanowire single-photon detector}

\newacronym[plural=ECDLs]{ecdl}{ECDL}{external cavity diode laser}

\newacronym{sfwm}{SFWM}{spontaneous four-wave mixing}

\newacronym[plural=EOMs]{eom}{EOM}{electro-optic modulator}

\newacronym{cw}{CW}{continuous-wave}

\newacronym[plural=SPADs]{spad}{SPAD}{single-photon avalanche diode}

\newacronym{hbt}{HBT}{Hanbury Brown-Twiss}
\newacronym{mrr}{MRR}{micoring resonator}

\newacronym{app}{$P_{\text{ap}}$}{afterpulsing probability}

\newacronym{dcr}{DCR}{dark count rate}

\newacronym{cr}{CR}{capacitance response}

\newacronym{pde}{PDE}{photon detection efficiency}

\newacronym{pcb}{PCB}{printed circuit board}

\newacronym{fwhm}{FWHM}{full width at half-maximum}

\newacronym{hsps}{HSPS}{heralded single-photon source}

\newacronym{rf}{RF}{radio frequency}

\newacronym{itu}{ITU}{International Telecommunication Union}

\newacronym{dwdm}{DWDM}{dense wavelength-division multiplexing}

\newacronym{fsr}{FSR}{free spectral range}

\newacronym{cmos}{CMOS}{complementary metal-oxide semiconductor}

\newacronym{pdh}{PDH}{Pound-Drever-Hall}

\newacronym{mfd}{MFD}{mode field diameter}

\newacronym{snr}{SNR}{signal-to-noise ratio}

\newacronym{g2}{$g^2(\tau)$}{second-order Glauber correlation}

\newacronym{pgr}{PGR}{pair generation rate}

\newacronym{car}{CAR}{coincidences-to-accidentals ratio}

\newacronym{nir}{NIR}{Near Infra Red}
\newcommand{\acro}[1]{\gls*{#1}}
\newcommand{\acropl}[1]{\glspl*{#1}}

\makeatletter
\def\@email#1#2{%
 \endgroup
 \patchcmd{\titleblock@produce}
  {\frontmatter@RRAPformat}
  {\frontmatter@RRAPformat{\produce@RRAP{*#1\href{mailto:#2}{#2}}}\frontmatter@RRAPformat}
  {}{}
}%
\makeatother
\begin{document}

\preprint{AIP/123-QED}

\title[Integrated Telecom Wavelength Heralded Single-Photon Source based on GHz gated detectors]{Integrated Telecom Wavelength Heralded Single-Photon Source based on GHz gated detectors}
\author{M.~A.~Pereira}
\altaffiliation[]{These authors contributed equally}
\author{M.~Wu}%
\altaffiliation[]{These authors contributed equally}
\affiliation{Department of Applied Physics, University of Geneva, Geneva, Switzerland}
\author{A.~S.~Raja}
\author{R.~N.~Wang}
\author{T.~Kippenberg}
\affiliation{Laboratory of Photonics and Quantum Measurements, EPFL, Lausanne, Switzerland}
\author{H.~Zbinden}
\affiliation{Department of Applied Physics, University of Geneva, Geneva, Switzerland}
 \affiliation{Vigo Quantum Communication Center, Universidade de Vigo, Vigo E-36310, Spain}
\author{T.~Brydges}
\author{R.~Thew}
\email{robert.thew@unige.ch}
\affiliation{Department of Applied Physics, University of Geneva, Geneva, Switzerland}

\date{\today}

\begin{abstract}
We introduce a simple and flexible concept for a heralded -- spectrally pure -- single photon source. The scheme uses a probabilistic photon pair source pumped with a CW laser, whereby a rapid gating InGaAs/InP single photon avalanche diode provides a synchronous clock and temporally resolves, and hence spectrally filters, the heralded photons. We demonstrate the concept by combining this with a narrow-band integrated silicon nitride photon-pair source. This simple architecture is capable of heralding photons with high spectral purity in the telecom band, but could be adapted to other wavelengths and bandwidth regimes.
\end{abstract}

\maketitle



\begin{quotation}
At the heart of all quantum photonic applications are technologies to generate and detect photons~\cite{Eisaman:2011}. Historically, this has been dominated by probabilistic sources based on spontaneous parametric down conversion (SPDC), to generate entangled photons, and semiconductor single photon avalanche detectors (SPAD) -- silicon for the visible regime and InGaAs/InP for the telecom. In the last few years, this hegemony has been challenged by progress on the one hand in quantum dots~\cite{Tomm:2021, Ding:2025}, whose performance is now surpassing that of probabilistic-based approaches such as SPDC, and on the other hand, the enormous advances in the performance of superconducting nanowire single photon detectors (SNSPD)~\cite{Morozov:2021}.

 Nonetheless, SPDC schemes and, more recently, spontaneous four wave mixing (SFWM), can be readily engineered for a wide variety of wavelengths and bandwidths, and can thus be adapted to suit a more diverse range of application scenarios. Along with SPADs, they also have the advantage of operating at room temperature, or with simple electrical temperature control. SPADs can also be operated in different detection regimes -- synchronous "gated", or asynchronous "free-running" -- as required. A particularly interesting gated-mode regime in the telecom (InGaAs/InP) band involves rapidly gating the SPAD with short, sub-ns, gates. This has advantages in terms of reduced noise (per gate) and afterpulsing, as well as having a lower detection jitter~\cite{Zhang:2010}. 

SPDC and SFWM produce time correlated photon pairs, whereby, the detection of one photon of the pair can be used to herald the other, giving rise to the name heralded single photon source (HSPS).  Variations of HSPS using synchronous or asynchronous gated detectors have been demonstrated, where synchronized schemes have also relied on pulsed or modulated pump lasers, often adding to the complexity and cost of the scheme. Typically, these have only produced pure (spectrally uncorrelated) photons with careful phasematching and pump bandwidth engineering~\cite{Mosley:2008}. 

In this work, a SFWM source is pumped with a \acro{cw} laser and, as such, the photon pairs can be generated at any time. However, as the detector's temporal resolution is much shorter than the photons' coherence time, the heralding detection projects the heralded photon into a well-defined, and now synchronously clocked, stream of photons. This simple approach to HSPS can be easily adapted to other photonic sources, provided that the photon coherence length is longer than the jitter of the detector~\cite{Huang:2010}. Indeed, the concept has previously been exploited to clock entanglement swapping experiments~\cite{Halder:2007,Samara:2021}. In the following we first explain how the SPAD functions, then describe the photon pair source, showing their characterizations before elaborating on the HSPS concept and presenting the results.
\end{quotation}


\acro{nir} free-running single photon detectors are essential for a number of applications, such as LiDAR and biomedical sensing \cite{Yu2018-uk,Liang2022-sx,Li2021-am,Slenders2021-ay,Bruschini2019-et}. Among the leading single photon detector technologies are \acro{sns} that offer exceptional performance with near-unity detection efficiency and ultra low dark counts. However, \acro{sns} rely on cryogenic cooling, which significantly increases their complexity and limits their suitability for some applications.
By contrast, InGaAs/InP \acro{spad} are widely used for \acro{nir} detection due to their simplicity, low cost, low power consumption, and small footprint. For high-speed performance, gated-mode \acro{spad}s are preferred -- often referred to as rapid gating. The small active time of the gated signal ensures low timing jitters as well as lower dark count rates and afterpulsing effects when compared to free-running or standard gated \acro{spad}s\cite{Namekata2010-kz,Patel2012-ac,Zhang:2010,Liang2011-oa, Tosi2013-ed}. 
The afterpulse probability $(P_{\rm ap})$ is a limiting factor for the maximum count rates in \acro{spad}s, as it generally requires long hold-off times (time after an avalanche event during which the detector is kept below its breakdown voltage) to be suppressed.

When InGaAs/InP \acro{spad}s are used in gated mode, the output signal of a typical \acro{spad} includes the avalanche signal (a few millivolts) and a strong parasitic signal that can be orders of magnitude higher than the avalanche signal. This parasitic signal is a \acro{cr} caused by the rapid change in voltage across the \acro{spad} when a gate is applied and is proportional to both the capacitance and the rate of voltage change. There are two obvious ways of detecting single photon events in these conditions, either the bias of the detectors is increased in order to have larger avalanches that can be easily discriminated from the \acro{cr} - resulting in higher \acro{dcr} and \acro{app} - or by using complex readout electronics to reduce the \acro{cr}. 
The first approach of increasing the bias voltage creates a problematic feedback loop: higher bias leads to more intense avalanche events, which trap more charge carriers in defects, increasing \acro{app}. To mitigate this, longer hold-off times are required, which in turn limits the maximum achievable count rate.

An alternative approach to this takes a copy of the signal, delays it by one period, and then subtracts the signals to eliminate this \acro{cr}~\cite{Yuan2007}. An extension of this concept introduced a second SPAD to provide a reference signal that could be instead subtracted~\cite{Scarcella2015-eq}. Park {\it et al.}~\cite{Park:19} took this a step further and developed a dual anode InGaAs/InP \acro{spad}s (DA-SPADs) that is comprised of two on-chip diodes separated by an isolation wall and sharing a common cathode. Both diodes are designed to operate as \acro{spad}s; however, one diode serves as a \textit{dummy}, and is engineered to have a higher breakdown voltage compared to the other diode, which makes it unable to detect single photons. In principle, the \textit{dummy} detector creates a \acro{cr} similar to that of the main detector, which is later removed from the detection's output signal via a subtraction circuit. This enables the detection of smaller avalanche signals that would otherwise be hidden by the \acro{cr}, allowing for a better discrimination and read-out efficiency of the detection signal, while  keeping \acro{dcr} and \acro{app} relatively low.

\begin{figure}
    \centering
    \includegraphics[width=1\linewidth]{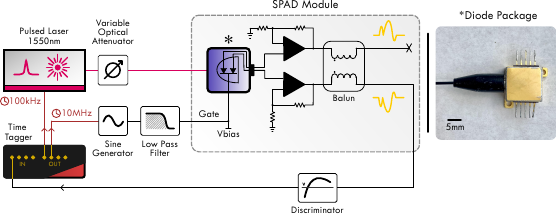}
    \caption{Schematic of the setup used for characterising the SPAD, along with a picture of the diode package.}
    \label{fig:charac_spadpackage}
\end{figure}
Fig.\ref{fig:charac_spadpackage} is a schematic representation of the experimental setup used for characterizing the detector's [WOORIRO, SPAD with internal TEC] \acro{pde}, \acro{dcr}, and \acro{app} using an attenuated laser. A pulsed laser at \SI{1550}{nm} was connected to a variable optical attenuator, reducing the mean photon number per pulse to 0.5. The electrical part of the experimental setup comprises a sine generator, used for the gate generation, sent to an RF amplifier followed by a low-pass filter, before being sent to the detector. A source-meter was used to supply the required DC voltage to the detector which was subsequently connected to a subtracting circuit. External to the detector's control PCB, the output signal was then discriminated, and the TTL signal sent to a time tagger [ID Quantique, ID900]. The time tagger also acted as a reference clock for the characterization setup, providing a \SI{10}{MHz} clock signal to the sine generator and a periodic trigger signal to a pulsed laser used for characterization. The detector was gated at $1$\,GHz, using a sine wave with $20\mathrm{V}_{\text{pp}}$, and an effective gate width below $\SI{300}{\pico\second}$.

To accurately measure the PDE, the laser repetition rate ($f_\text{L}$) was set to \SI{100}{kHz} and synchronized with the detector gating. This low repetition rate provides enough time between laser pulses preventing afterpulsing events from artificially inflating the PDE. The characterization of \acro{pde}, \acro{dcr}, and \acro{app} was then performed in a novel approach by collecting a single histogram spanning the totality of the pulsed laser's period ($\SI{10}{\micro\s}$), with bin-widths as to include one gate/bin (\SI{1}{ns}), corresponding to the \SI{1}{GHz} gate frequency.

\begin{figure}[!ht]
        \centering
        \includegraphics[width=0.48\textwidth]{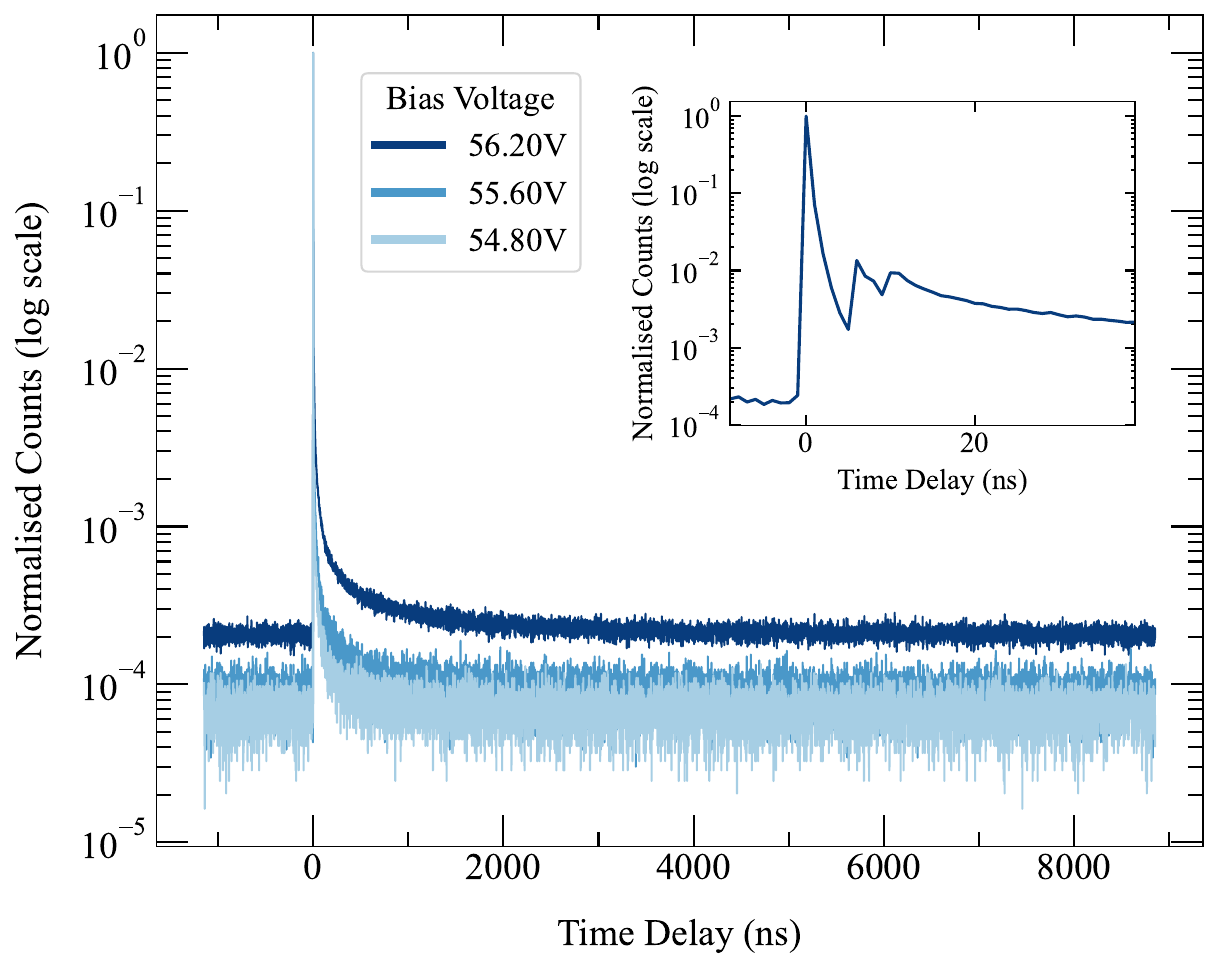}
        \caption{Detection count histograms (normalized to the maximum) showing the time delay distribution of after-pulsing and dark count events relative to laser pulse events, which is defined to be at \SI{0}{ns} delay. Data collected at different SPAD bias voltages are shown to illustrate the increase in afterpulsing occurrence and dark count rate with bias voltage. Inset: Data for a bias voltage of \SI{56.20}{V}, shown near \SI{0}{ns}. The integration time was \SI{120}{s}.}
        \label{fig:afterpulse_plot}
    \end{figure}

Traditionally, the PDE is calculated using the expression

\begin{equation}
\text{PDE}=\frac{1}{\mu} \ln \left(\frac{1-P_\text{d}}{1-P_\text{t}}\right)\,,
\label{old_pde_equation}
\end{equation}
where $P_\text{d}$ and $P_\text{t}$ are the dark count probability and total count probability respectively, and $\mu$ is the mean photon number per laser pulse. The natural logarithmic terms are Poissonian corrections when converting between detection probabilities and detection rates~\cite{lu2011}. Since our setup allowed access to all the key parameters directly, we used an alternative method to calculate the PDE. Figure~\ref{fig:afterpulse_plot} depicts sample data representing the temporal distribution of afterpulsing/dark count events occurring after each laser pulse. The characterization used short laser pulses, <\SI{50}{ps} full width at half maximum (FWHM), that were aligned in time such that they were completely contained in a single histogram bin, with bin-width $\Delta t = \SI{1}{\nano\s}$. As seen in the plot inset, after the initial laser irradiation peak, a sharp decrease in counts in the subsequent \SI{10}{ns} was observed. This behaviour emerges from the discriminator's inherent deadtime, caused by the bandwidth limitations of its monostable circuit (see Supplementary Material).

The PDE can then be directly calculated from the signal counts in the time bin corresponding to the arrival of the laser pulse, while correcting for dark counts:
\begin{equation}
\text{PDE} = \frac{C_\text{L} - DCR\times\Delta t}{\mu' \times N_\text{L}}\,, 
    \label{pde_equation}
\end{equation}
where $C_\text{L}$ represents the total counts in the signal bin, $\Delta t$ the histogram's bin width and $N_\text{L}$ is the total number of pulses. A corrected mean photon number per pulse, $\mu'$, is defined by 

\begin{equation}
\mu' = 1 - \frac{\mu^{n}e^{-\mu}}{n!}\Bigg\vert_{n=0}= 1-e^{-\mu}\,.
    \label{mu_equation}
\end{equation}
This represents the Poissonian correction to the mean photon number ($\mu$), calculated using an average optical power measurement to obtain the probability of having \textit{at least} one photon per pulse. Finally, the \acro{dcr} can be calculated either by measuring the histogram with the laser off, or by only considering the counts at very large time delays.

The \acro{app} can be characterized by physically implementing different hold-off times in the discrimination circuit. However, this would require multiple dataset measurements and also requires physical modifications to the discriminator hardware between each measurement. Here, the \acro{app} was calculated for various arbitrarily set hold-off times ($t_\text{d}$) through post-processing, by discarding the counts in the $t_\text{d}$ bins following the laser illuminated bin from the histogram.

The \acro{app} is then given by the ratio of counts due to afterpulsing (total non-illuminated counts collected on histogram) and illuminated counts, both corrected for \acro{dcr}:
\begin{equation}
P_{\text{ap}} = \frac{C_\text{T}  - C_\text{L} - DCR\times \Delta T}{C_\text{L}-DCR \times \Delta t}\,,
    \label{app_equation}
\end{equation}
where $C_\text{T}$ refers to the total counts for a chosen $t_\text{d}$, $\Delta T$ is the corresponding integration time, and $C_\text{L}$ is the total counts in the signal bin. Importantly, this approach achieves the same result as physically implementing different hold-off times in the discriminator circuit, but requires only one histogram for all analyses and eliminates the need to modify the discriminator hardware.

\begin{figure}[t]
        \centering
        \includegraphics[width=0.48\textwidth]{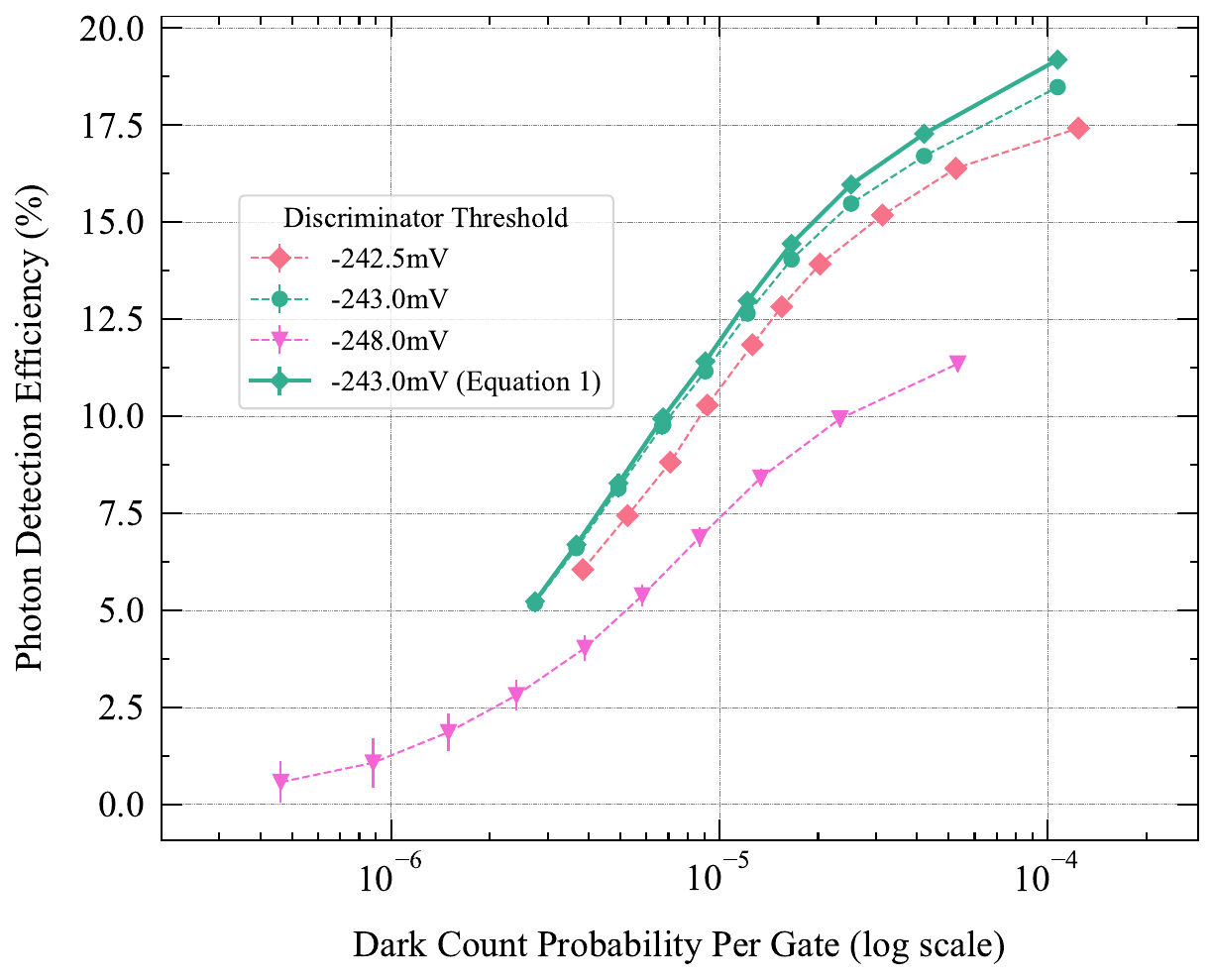}
        \caption{Characterization data of the SPAD detector, where corresponding dark count probability per gate and photon detection efficiencies (PDE) are shown against varying discriminator threshold values. Dashed lines show the PDE calculated from Eq.~\ref{pde_equation}, with the solid line showing the PDE calculated from Eq.~\ref{old_pde_equation}. The PDE is adjusted by changing the SPAD bias voltage.}
        \label{fig:pde_dcr}
    \end{figure}

The discriminator threshold level, $V_{\text{th}}$, can have a profound effect on the performance of the detector. $V_{\text{th}}$ should be set as low as possible to catch all small avalanches from photon events to improve the signal readout efficiency. However, due to imperfections in the detector and associated electronic components, the measured signal is distorted in the gates before and after an avalanche signal. As such, a compromise has to be made on the efficiency to avoid picking up these noisy readout signals. Fig.~\ref{fig:pde_dcr} shows the calculated efficiency from Eq.~\ref{pde_equation} as a function of dark count probability per gate for various discrimination threshold settings, where we obtained the best PDE to noise performance with $V_{\text{th}}=-243.0\,\mathrm{mV}$, as well as settings with higher and lower voltage thresholds to illustrate their reduced performance. A corresponding curve using Eq.~\ref{old_pde_equation} to calculate the PDE is also shown for comparison: We saw generally good agreement, with Equation \ref{old_pde_equation} giving a slightly higher PDE. While this does not affect the performance of the HSPS characteristics, further work is required to better understand this small discrepancy. For the following measurements, we chose a \acro{pde} working point of 15.5\,\%, which was the maximum PDE before the rate of increase of the \acro{dcr} became more disproportionate to that of the PDE.  

Finally, the jitter was characterized from measurements obtained using a similar time-tagging setup in Fig.~\ref{fig:charac_spadpackage}. Under the same PDE conditions, we measured the upper bound for the FWHM jitter to be \SI{30}{ps}, significantly shorter than for typical \acro{spad}s, where jitter usually exceeds \SI{100}{ps}~\cite{Amri:16, farrell}. Measurement data and setup details may be found in the Supplementary Material.

\begin{figure}[t]
        \centering
        \includegraphics[width=0.49\textwidth]{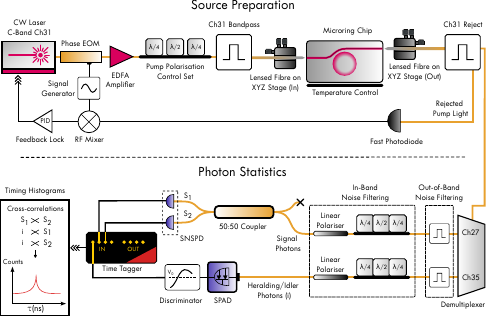}
        \caption{Schematic of the experimental setup, starting with the photon-pair source preparation stage followed by the photon statistics stage. The former concerns the generation of photon pairs by coupling laser light into the microring and locking the pump frequency to a resonant mode of the microring. The latter filters out residual pump light and frequency-demultiplexes the photon pairs for detection in a (heralded) \acro{hbt} setup.}
        \label{fig:setup2}
    \end{figure}

A key component of the fully fibre-based \acro{hsps} presented here was the microring resonator photon pair source, based on a silicon nitride integrated photonic chip. Telecom-wavelength photon-pairs (signal-idler) were generated via SFWM. These pairs were out-coupled and frequency demultiplexed using \acro{dwdm} filters, with the idler photons detected as heralds for the signal photons. The integrated fibre-based design allowed the setup to be highly compact and robust. Fig.~\ref{fig:setup2} shows a schematic of the full experimental setup. The MRR had a 200GHz free spectral range and was pumped at \SI{1552.5}{nm} with light from a \acro{cw} laser. For further details on the source see the Supplementary Material.

The MRR chip temperature was actively stabilized, and the \acro{pdh} technique~\cite{Drever:1983} was used to actively lock the laser frequency to the cavity resonance using a Red Pitaya and home-built amplifier board. The device had Q-factors for pump, signal, and idler on the order of $3.5 \times 10^{6}$, corresponding to signal (idler) bandwidths of about \SI{53}{\mega\hertz} (\SI{60}{\mega\hertz}).

\begin{figure}[t]
        \centering
        \includegraphics[width=0.49\textwidth]{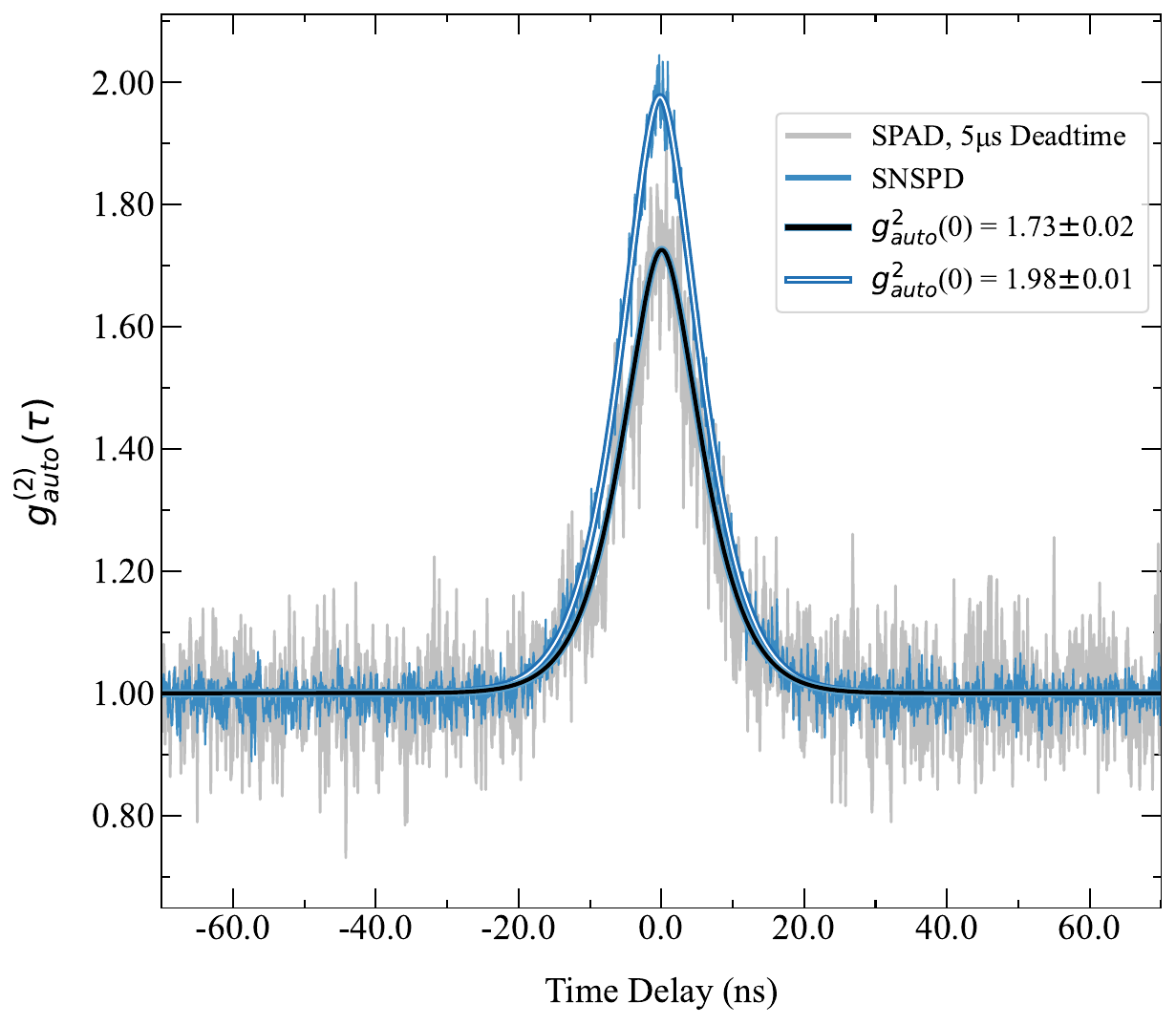}
        \caption{Auto-correlation measurement of the signal photon stream using an SNSPD and either our SPAD or another SNSPD, as well as the data fits. The SPAD had a software-level \SI{5}{\micro s} deadtime applied to the timetags to filter afterpulsing events.}
        \label{fig:selfcorr_deadtime}
    \end{figure}

    \begin{table}[b]
        \caption{\label{tab:table1}Detector parameters for characterising the HSPS, with fixed gating frequency of \SI{1}{GHz}. $P_{\text{DC}}$ is the dark count probability per gate.}
        \begin{ruledtabular}
        \begin{tabular}{cccccc}
        PDE(\%) & $P_{\text{DC}}$ & $V_{\text{th}}$(mV) & $t_\text{d}$($\SI{}{\micro\second}$) & $P_{\text{ap}}$(\%) & Gate Width(ps)\\
        \hline
         15.5 & $1.25\times10^{-5}$ & -243 & 5 & $<1.0$ & <300\\
        \end{tabular}
        \end{ruledtabular}
    \end{table}
To quantify the performance of the HSPS, the purity of the source was measured with the SPAD detecting the heralding photon, and an SNSPD detecting the heralded photon, with the experimental setup as shown in Fig.~\ref{fig:setup2}. The SPAD settings are given in Table~\ref{tab:table1} and photon-pair source parameters given in Table~\ref{tab:table2}. The unheralded second-order correlation function, $g^{(2)}_{auto}(\tau)$, is shown in Fig.~\ref{fig:selfcorr_deadtime}. The $g^{(2)}_{auto}(0)$ can be straightforwardly related to the photon purity via $P = g^{(2)}_{auto}(0)-1$, with the HSPS having a purity of $P=0.73\pm0.02$. For comparison, we obtained $g^{(2)}_{auto}(0) = 1.98\pm0.01$ ($P=0.98\pm0.01$) using two SNSPDs instead (Fig.~S4~a, Supplementary Material), indicating the MRR itself produces photon pairs with very high photon purity.

The shortfall in purity when using our SPAD can be mainly attributed to a higher \acro{dcr} as the $g^{(2)}_{auto}(0)$ is not dependent on the detector efficiency, and the jitter of the two types of detectors is comparable. Depending on the application, one can choose a lower $V_{\text{bias}}$ setting to reduce the \acro{dcr} at the expense of \acro{pde}. To characterize the single-photon nature of the HSPS, the heralded autocorrelation, $g^{(2)}_{h}(0)$ was measured. A value of $0.198\pm0.005$ was obtained for an MRR chip input power of \SI{235}{\micro\watt}, an integration time of \SI{7752}{\second}, and a coincidence window of \SI{3}{\nano\second}. This pump power corresponds to a photon pair probability of $\sim0.2\%$ per gate and $\sim3\%$ per coherence time.

    \begin{table}[b]
        \caption{\label{tab:table2}Photon-pair source parameters for characterising the HSPS, with the pump power at chip input being $\SI{660}{\micro\watt}$. $\eta_{\text{coup}}$ is the average coupling efficiency from source to detector, with PGR the Pair Generation Rate (see Supplementary Material).}
        \begin{ruledtabular}
        \begin{tabular}{cccc}
        $\Delta\nu_{s}\,[\Delta\nu_{i}$]\,\,(MHz) & $\eta_{\text{coup}}$(\%) & PGR(MHz) & $g^{(2)}_{auto}(0)$  \\
        \hline
        52.8\,[59.8] & $\sim$44 & $\sim7.5$ & 1.98(1)  \\
        \end{tabular}
        \end{ruledtabular}
    \end{table}
An important quantity to characterise the performance of a HSPS is the heralding efficiency - the probability of the heralded photon being in the output mode, $\eta_{\text{h,s}}$. To calculate this, the coincidence count rate $R_{\text{s,i}}$ within a coherence time of the signal and idler photons was first determined from:
\begin{equation}
R_{\text{s,i}} = \frac{1}{\Delta T} \sum\limits_{\tau>-\tau_{\text{c,s}}}^{\tau<\tau_{\text{c,i}}}C_{\tau}  ,
    \label{coincount_equation}
\end{equation}
where the summation represents the total counts from time bins where $\tau$ is within the (asymmetric) photon coherence times $\tau_\text{{c,s}}$ and $\tau_\text{{c,i}}$ and $\Delta T$ is the integration time. $\tau_\text{{c,s}}$ and $\tau_{c,i}$ are extracted from the signal-idler cross-correlation histogram (see Supplementary Material). 

The signal photon heralding efficiency, $\eta_{\text{h,s}}$, can then be calculated from: 
\begin{equation}
\eta_{\text{h,s}} = \frac{R_{\text{s,i}}}{R_{\text{i}} \eta_{\text{d,s}}}\, ,
    \label{h_eff_equation}
\end{equation}
where $\eta_{\text{d,s}}$ is the heralded (signal) detector \acro{pde} and $R_\text{i}$ is the idler photon count rate measured on the SPAD \cite{signorini}. The heralded signal photon count rate ($R_{\text{h,s}}$) follows as $\eta_{\text{h,s}}R_{\text{i}}$. 

\begin{figure}[t]
        \centering
        \includegraphics[width=0.49\textwidth]{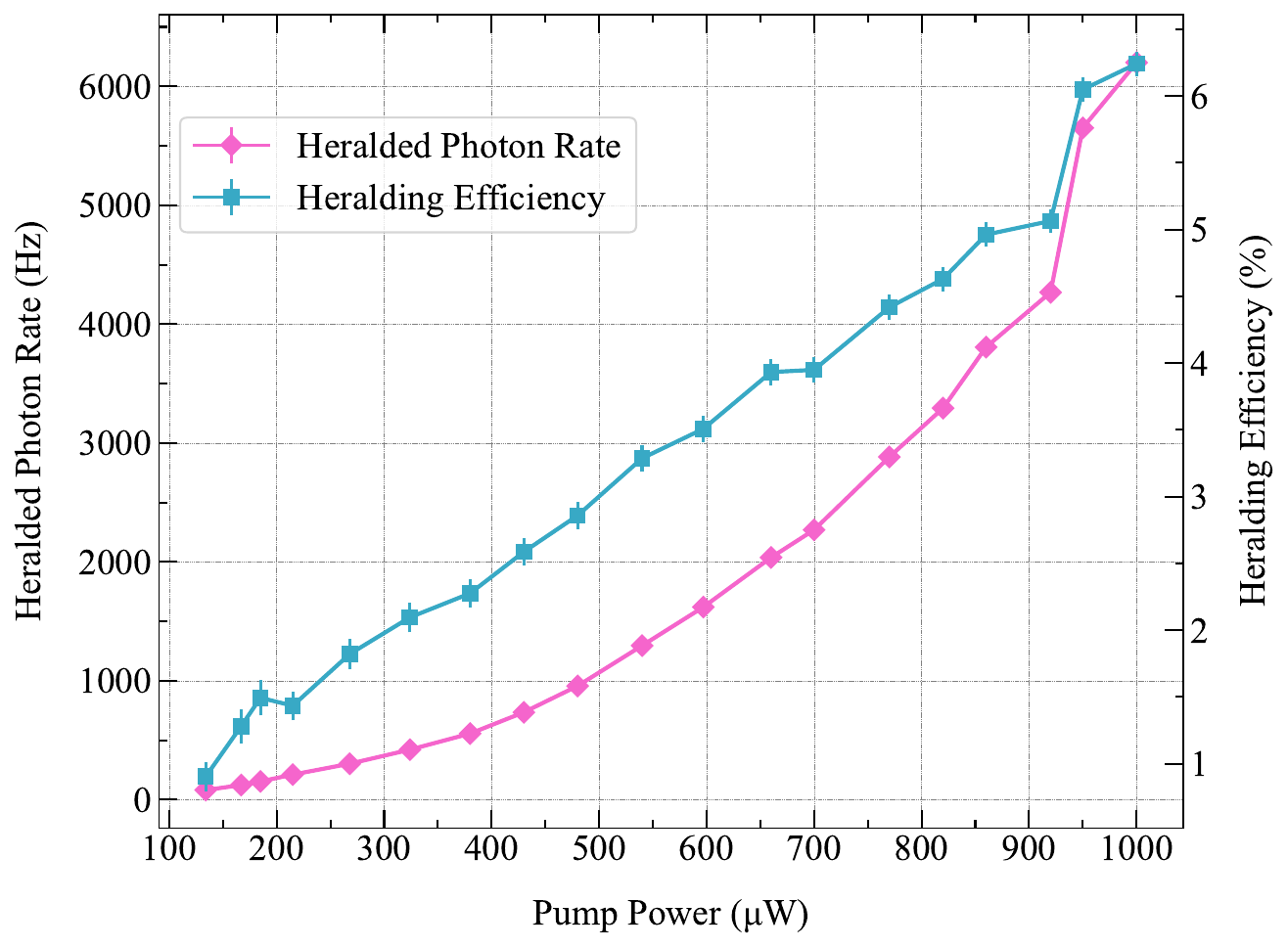}
        \caption{The heralded signal photon rate and heralding efficiency as functions of pump power. Error bars are derived from the Poissonian error of the photon count rates.}
        \label{fig:pgr}
    \end{figure}

Fig.~\ref{fig:pgr} shows $\eta_{\text{h,s}}$ and the photon heralding rate for the 53\,MHz bandwidth signal photons as functions of pump power. We observed the photon heralding efficiency increasing with pump power, which we can ascribe to the relatively high \acro{dcr} of the \acro{spad}: As the photon pair rate increases with a constant \acro{dcr}, the probability of detecting a true (idler photon) heralding event per gate also increases. On the other hand, the heralding efficiency is adversely affected by transmission losses through the channel between the MRR chip and detector, however, these losses can be improved with more optimized filtering. In this proof of concept demonstration we used $\SI{660}{\micro\watt}$ of pump power at the MRR chip, observing $\eta_{\text{h,s}}\sim3.9\%$ and $R_{\text{h,s}}\sim\SI{2.0}{\kilo\hertz}$ for photons with bandwidths around \SI{50}{\mega\hertz}.

This paper has shown the proof-of-principle demonstration of a telecom \acro{hsps} comprised of an integrated photon pair source in combination with a dual-anode InGaAs/InP SPAD \cite{Park:19} operated at GHz gate rates. Due to the CW pumping of the source, the SPAD is not only used for single-photon detection, but is also used as the synchronization signal for clocking the entire system. In this work the SPAD was operated at a \SI{1}{GHz} rate, however, detectors suitable for this scheme can be operated at much higher rates, in excess of \SI{2}{GHz} \cite{Zhang:2010, Patel2012-ac}. Although the demonstration in this work used a very narrowband source, the SPAD's low jitter would allow its use with photon bandwidths above even \SI{20}{GHz}, while still allowing the detection of the heralding photon to project the heralded photon into a well-defined, synchronously clocked stream of photons \cite{Huang:2010}, depending on the desired application.

Our \acro{hsps} setup is currently limited by the noise characteristics of the SPAD. This could be addressed through developing a more advanced detector control PCB, to allow for adjustment of both phase and amplitude of each diode's \acro{cr} within the subtraction circuit, which would suppress the background \acro{cr} signal further and thus allow a better readout efficiency with substantially reduced \acro{dcr} and \acro{app} for the same PDE. Ideally, though, this should also be addressed at the fabrication stage for the dual-anode InGaAs/InP SPADs. The simple nature of the concept and all \acro{hsps} components in this work demonstrates the potential for future, improved \acropl{hsps} in field-deployed quantum networks.\\

The authors thank Alberto Boaron and ID Quantique for the dual-anode InGaAs/InP SPAD. We acknowledge David Cabrerizo for assistance with the in-house electronics development. This work was supported by the Swiss State Secretariat for Research and Innovation (SERI) (Contract No. UeM019-3). T.B. and M.W. are supported by the Swiss National Science Foundation through Ambizione Grant No. PZ00P2$\_$216153.

\section*{Author Declarations}

\subsection*{Conflict of Interest}
The authors declare no conflicts of interest.

\section*{Data Availability}

Data available on request from the authors.

\bibliography{Bibliography}

\end{document}